\documentclass[twocolumn,A4]{article}
\usepackage[dvips]{graphics}
\usepackage{setspace}
\usepackage{color}
\definecolor{gold}{rgb}{0.85,0.66,0}
\definecolor{dblue}{rgb}{0,0,0.8}

\begin{document}

\onecolumn

\begin{center}
{\bf{\Large {\textcolor{gold}{Magnetic Response in Mesoscopic Rings and 
Moebius Strips: A Theoretical Study}}}}\\
~\\
{\textcolor{dblue}{Santanu K. Maiti}}$^{\dag,\ddag}$,\footnote{{\bf
Corresponding Author}: Santanu K. Maiti \\
$~$\hspace {0.45cm} Electronic mail: santanu.maiti@saha.ac.in} \\
~\\
{\em $^{\dag}$Theoretical Condensed Matter Physics Division,
Saha Institute of Nuclear Physics, \\
1/AF, Bidhannagar, Kolkata-700 064, India \\
$^{\ddag}$Department of Physics, Narasinha Dutt College,
129 Belilious Road, Howrah-711 101, India} \\
~\\
{\bf Abstract}
\end{center}
We investigate magnetic response in mesoscopic rings and moebius strips 
penetrated by magnetic flux $\phi$. Based on a simple tight-binding
framework all the calculations are performed numerically which describe
persistent current and low-field magnetic susceptibility as functions
of magnetic flux $\phi$, total number of electrons $N_e$, system size
$N$ and disorder strength $W$. Our exact analysis may provide some 
important signatures to study magnetic response in nano-scale loop 
geometries.
\vskip 1cm
\begin{flushleft}
{\bf PACS No.}: 73.22.-f; 73.23.-b; 73.23.Ra; 75.20.-g \\
~\\
{\bf Keywords}: Mesoscopic ring; Moebius strip; Persistent current; 
Low-field magnetic susceptibility; Disorder.
\end{flushleft}

\newpage
\twocolumn

\section{{\textsl{Introduction}}}

More advanced nanoscience and technology has made it possible to fabricate
devices whose dimensions are comparable to mean free path of an electron
and electron transport in such systems gives several novel and interesting 
new phenomena. In the mesoscopic regime, phase coherence of the electronic 
states are of fundamental importance, and the phenomenon of persistent 
current is a spectacular consequence of quantum phase coherence in this 
regime. Experimental investigations on these systems have provided several 
surprising quantum behaviors in contrast to those anticipated from the 
classical theory of metals. At much low temperatures, two aspects of the 
new quantum regime appear that are of particular interest. \\
\noindent
(i) The phase coherence length $L_{\phi}$, the length scale over which an 
electron maintains its phase memory, increases significantly with the 
decrease of temperature and is comparable to the system size $L$. \\
\noindent
(ii) The energy levels of these small finite size systems are discrete. \\
\noindent
These two are the essential criteria for the existence of persistent 
current in these systems upon the application of an external magnetic 
flux $\phi$. Many theoretical works~\cite{byers,butt,cheu1,cheu2,mont,
alts,von,schm,ambe,bouz,giam,yu,kulik,avishai,weiden,san1,san2,san3} 
on persistent current 
in one-dimensional ($1$D) mesoscopic rings are available in literature, 
yet they cannot clearly explain several results those have been obtained 
experimentally. The discovery of persistent current in mesoscopic rings 
has addressed new interesting questions on the thermodynamics of these 
systems. Although such an effect was predicted several years ago, the 
unexpectedly large amplitudes of measured currents lead to many important
questions. It has been proposed that the electron-electron (e-e) 
interactions contribute significantly to the average currents. Although, 
an explanation based on the perturbative calculation both in interaction 
and disorder, seems to give a quantitative estimate closer to the 
experimental results but still it is less than the measured currents 
by an order of magnitude, and the interaction parameter used in the 
theory is not well understood physically. Though the enhancement of 
current amplitude can be understood by some theoretical 
arguments~\cite{san1,san2,san3} but the sign of the currents cannot 
be predicted precisely. Levy {\em et al.}~\cite{levy} have observed 
diamagnetic response for the measured currents at low fields in the 
experiment on $10^7$ isolated mesoscopic Cu rings. While, Chandrasekhar 
{\em et al.}~\cite{chand} have measured $\phi_0$ periodic currents in 
Ag rings with paramagnetic response near zero fields. In a theoretical 
work Cheung {\em et al.}~\cite{cheu2} have predicted that the direction 
of current is random depending on the total number of electrons in the 
system and the specific realization of the random potentials. On the 
other hand, Yu and Fowler~\cite{yu} have shown both diamagnetic and 
paramagnetic responses in mesoscopic Hubbard rings. In an experiment, 
Jariwala {\em et al.}~\cite{jari} have obtained diamagnetic persistent 
currents with both $h/e$ and $h/{2e}$ flux periodicities in an array of 
$30$-diffusive gold rings. Similar diamagnetic sign of the currents in 
the vicinity of zero magnetic field were also found in an 
experiment~\cite{deb} on $10^5$ disconnected Ag rings. Thus we can 
emphasize that the prediction of the sign of low-field currents is a 
major challenge in this area. It has been studied that only for 
single-channel rings, the sign of the low-field currents can be 
mentioned exactly~\cite{san4,san5}. While, in all other cases i.e., 
for multi-channel rings and cylinders, sign of the low-field currents 
cannot be predicted exactly. Hence we see that the sign of low-field
currents is not very clear and the theoretical and  experimental results 
still do not agree very well. Therefore, we can say that the phenomenon 
of persistent current in metallic systems gives an exciting curiosity 
and is an open subject.

The paper is organized as follows. Following the brief introduction 
(Section $1$), in Section $2$, we describe the behavior of persistent 
current in strictly one-channel mesoscopic rings. The effect of 
intra-chain interaction on persistent current in moebius strips is 
discussed in Section $3$. Section $4$ illustrates the behavior of 
low-field magnetic susceptibility both for one-channel mesoscopic
rings and moebius strips, and finally, we summarize our study in 
Section $5$.

\section{{\textsl{Persistent current in one-channel mesoscopic rings}}}

\begin{figure}[ht]
{\centering \resizebox*{5.5cm}{3.5cm}{\includegraphics{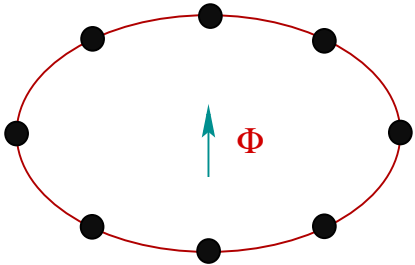}}\par}
\caption{{\textsl{Schematic view of a $1$D mesoscopic ring penetrated 
by a magnetic flux $\phi$. The filled black circles correspond to the 
position of atomic sites. A persistent current $I$ is established in 
the ring.}}}
\label{ring}
\end{figure}

Here we explore the behavior of persistent current in strictly one-channel 
mesoscopic rings. Let us start by referring to Fig.~\ref{ring} where a
mesoscopic ring is penetrated by a magnetic flux $\phi$.
To describe the system we use a tight-binding framework. Within a 
non-interacting electron picture, the tight-binding Hamiltonian for 
a $N$-site ring, enclosing a magnetic flux $\phi$ (in units of the 
elementary flux quantum $\phi_0=ch/e$), can be expressed in Wannier 
basis as,
\begin{equation}
H=\sum_i \epsilon_i c_i^{\dagger}c_i+\sum_{<ij>}v \left(e^{i\theta}
c_i^{\dagger}c_j + e^{-i\theta} c_j^{\dagger}c_i\right)
\label{hamil1}
\end{equation}
where, $\epsilon_i$'s are the on-site energies, $v$ is the hopping 
strength between nearest-neighbor atomic sites and $\theta=2\pi 
\phi/N$ is the phase factor due to the flux threaded by the ring. 
As the magnetic field associated with flux $\phi$ does not penetrate 
anywhere of the circumference of the ring, we neglect Zeeman term in 
the above Hamiltonian (Eq.~\ref{hamil1}). Throughout our calculations 
we set the hopping strength $v=-1$ and use the units where $c=e=h=1$.

In order to study current-flux ($I$-$\phi$) characteristics first it is 
necessary to know the variation of energy levels with $\phi$. For an
ordered ring, we calculate the energy eigenvalues analytically. 
Mathematically, the energy eigenvalue for $n$-th eigenstate can be 
written in the form,
\begin{equation}
E_n(\phi)=2v\cos\left[\frac{2\pi}{N}\left(n+\frac{\phi}
{\phi_0}\right)\right]
\end{equation}
where, $n$ is an integer bounded in the range $-N/2\leq n <N/2$. But
as long as impurities are introduced in the ring i.e., the ring becomes
a disordered one, energy eigenvalues cannot be obtained analytically.
In that case we do exact numerical diagonalization of the tight-binding
Hamiltonian to achieve the eigenvalues.
\begin{figure}[ht]
{\centering \resizebox*{7.5cm}{5.7cm}{\includegraphics{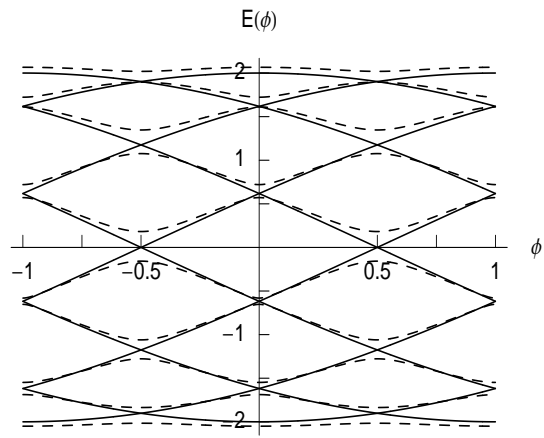}}\par}
\caption{{\textsl{Energy-flux characteristics for ordered (solid line) 
and disordered (dotted line) mesoscopic rings with $N=10$. For the 
disordered case, we choose $W=1$.}}}
\label{energy}
\end{figure}
As illustrative examples, in Fig.~\ref{energy} we plot the energy-flux
($E$-$\phi$) characteristics for a mesoscopic ring considering $N=10$.
The solid and dotted curves correspond to the perfect and disordered
rings, respectively. For the disordered ring, we set $W=1$. In absence
of any impurity ($W=0$), energy levels intersect with each other at
integer or half-integer multiples of $\phi_0/2$, while gaps open at
these points in presence of impurity. In both the two cases energy 
levels vary periodically as a function of $\phi$ showing $\phi_0$ 
($=1$, in our chosen unit $c=e=h=1$) flux-quantum periodicity.

The current $I_n$ carried by $n$-th energy level having energy $E_n$ is 
obtained by taking the $1$-st order derivative of $E_n$ with flux 
$\phi$. Mathematically it can expressed as,
\begin{equation}
I_n=-\frac{\partial E_n}{\partial \phi}
\end{equation} 
At absolute zero temperature ($T=0$K), total current of a system, 
described with fixed number of electrons $N_e$, will be obtained by
\begin{figure}[ht]
{\centering \resizebox*{7.5cm}{9.5cm}{\includegraphics{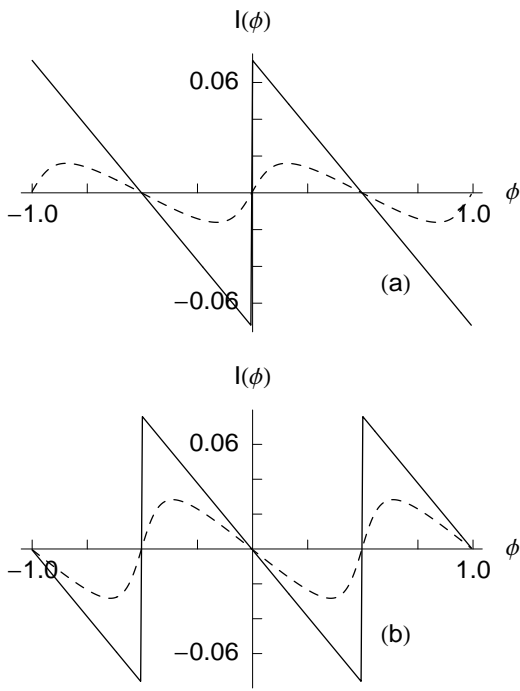}}\par}
\caption{{\textsl{Persistent current as a function of $\phi$ for ordered 
(solid line) and disordered (dotted line) mesoscopic rings with $N=150$, 
where (a) $N_e=50$ and (b) $N_e=55$. For the disordered case, we fix 
$W=1$.}}}
\label{ringcurr}
\end{figure}
taking the sum of individual contributions from the lowest $N_e$ energy
levels. Thus, we can write the total current in the form,
\begin{equation}
I(\phi)=\sum_n I_n(\phi)
\end{equation}
As representative examples, in Fig.~\ref{ringcurr} we show the variation
of persistent current for a typical mesoscopic ring considering $N=150$,
where (a) and (b) represent the results for even ($N_e=50$) and odd
($N_e=55$) number of electrons, respectively. The solid and dotted 
curves correspond to the perfect ($W=0$) and disordered ($W=1$) rings, 
respectively. For the ordered ring, we set $\epsilon_i=0$ for all $i$,
while for the disordered case site energies ($\epsilon_i$) are chosen
randomly from a ``Box" distribution function of width $W=1$. From the
results it is observed that, in impurity free rings current exhibits
saw-tooth like variation as a function of $\phi$ providing sharp 
transitions, associated with the crossing of energy levels, at 
$\phi=\pm n\phi_0$ or $\pm n\phi_0/2$, depending on whether the ring 
contains even or odd number of electrons. On the other hand, this
saw-tooth like behavior completely disappears as long as impurities 
are introduced in the ring. This is due to the fact that, in presence
of impurities all the degeneracies move out, and accordingly, energy 
levels vary continuously with $\phi$ which provide continuous variation 
of the current. Addition to this feature, it is also important to
note that in disordered ring current amplitude gets reduced 
significantly compared to the ordered one. Such a behavior can be 
well understood from the theory of Anderson localization, where we get 
more localization with the increase of disorder strength $W$~\cite{lee}. 
Both for perfect and dirty rings, persistent current varies periodically
with $\phi$ showing only $\phi_0$ flux-quantum periodicity.

\section{{\textsl{Persistent current in moebius strips}}}

\begin{figure}[ht]
{\centering \resizebox*{5cm}{4cm}{\includegraphics{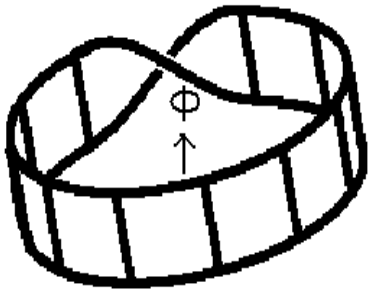}}\par}
\caption{{\textsl{Schematic representation of a moebius strip penetrated
by a magnetic flux $\phi$.}}}
\label{mobius}
\end{figure}

This section describes the behavior of persistent current in moebius
strips. Persistent current in a moebius strip is highly sensitive on 
the intra-chain interaction strength and here we will focus our study 
in this particular aspect. The schematic view of a moebius strip, 
penetrated by a magnetic flux $\phi$, is shown in Fig.~\ref{mobius}.
The vertical line connecting the atomic sites in upper and lower 
strands is called as `rung'. In presence of magnetic flux $\phi$,
the tight-binding Hamiltonian for a moebius strip with $N$ rungs 
(i.e., $2N$ atomic sites since in each strand there are $N$ atomic 
sites) can be written within the non-interacting electron picture as,
\begin{eqnarray}
H & = & \sum_{i=1}^{2N}\epsilon_i c_i^{\dagger}c_i + v\sum_{i=1}^{2N}
\left(e^{i\theta}c_i^{\dagger}c_{i+1} + e^{-i\theta}c_{i+1}^{\dagger}c_i 
\right) \nonumber \\ 
 & & + ~~v_{\bot}\sum_{i=1}^{2N} c_i^{\dagger}c_{i+N}
\end{eqnarray}
where, $\epsilon_i$ represents the on-site energy of an electron in 
site $i$ and $\theta=2\pi\phi/N$ is the phase factor associated with 
flux $\phi$. $v$ gives the nearest-neighbor hopping integral in each 
strand and $v_{\bot}$ corresponds to the hopping strength between the 
nearest-neighbor sites in the two strands, the so-called intra-chain
interaction strength. It ($v_{\bot}$) significantly controls the 
periodicity of persistent current in such a special type of geometric
model which we will describe in forthcoming sub-section.

\begin{figure}[ht]
{\centering \resizebox*{7.5cm}{5.5cm}{\includegraphics{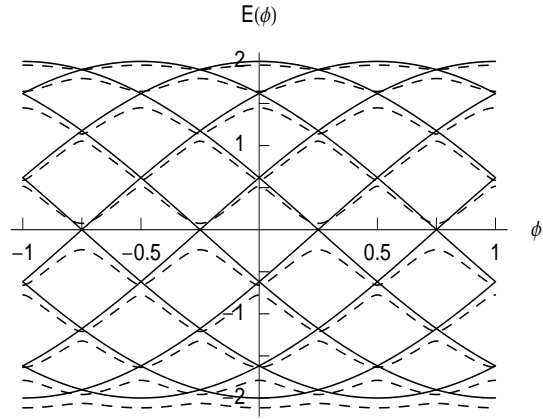}}\par}
\caption{{\textsl{Energy-flux characteristics for ordered (solid curve) 
and disordered (dotted curve) moebius strips ($N=5$) with $v_{\bot}=0$.}}}
\label{mobeng1}
\end{figure}

Now we start our discussion with the energy-flux characteristics of a
moebius strip. For illustrative examples, in Fig.~\ref{mobeng1} we
plot the energy levels as a function of $\phi$ for a moebius strip
considering $N=5$ in the typical case where $v_{\bot}=0$. The solid
and dotted lines correspond to the perfect and disordered moebius
strips, respectively. The presence (disordered strip) or absence 
(ordered strip) of $W$ provides exactly the similar behavior i.e.,
gap or gap less spectrum as we observe in the case of one-channel 
mesoscopic rings. Most importantly we notice that the energy levels 
have extremas at integer multiples of $\phi=\pm n\phi_0/4$ or 
$\pm n\phi_0/2$. Addition to this, it is also observed that the 
energy levels vary periodically
with flux $\phi$ providing $\phi_0/2$ flux-quantum periodicity instead 
of simple $\phi_0$ periodicity, as observed in traditional one-channel
mesoscopic rings or multi-channel cylinders. This phenomenon of half
flux-quantum periodicity can be explained as follow. In the typical
\begin{figure}[ht]
{\centering \resizebox*{7.5cm}{5.5cm}{\includegraphics{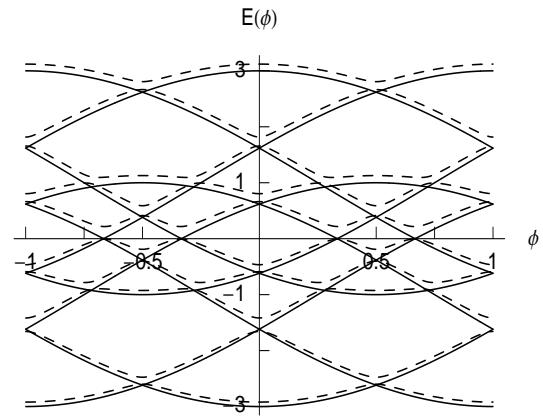}}\par}
\caption{{\textsl{Energy-flux characteristics for ordered (solid line) 
and disordered (dotted line) moebius strips ($N=5$) with $v_{\bot}=-1$.}}}
\label{mobeng2}
\end{figure}
case where $v_{\bot}=0$, electrons cannot able to hop along the
vertical direction. Therefore, in a moebius strip if an electron 
starts to move from any point, it will come back at its initial 
position after traversing twice the path length, and accordingly,
it encloses $2\phi$ flux. This provides $\phi_0/2$ flux-quantum 
periodicity and it is the speciality of such a twisted geometry.
Now, as long as electrons are allowed to hop between the two strands 
i.e., $v_{\bot}\ne 0$, energy levels get back $\phi_0$ flux-quantum 
periodicity. This is quite obvious since for the case when $v_{\bot}\ne 
0$ one can treat the moebius strip as a conventional cylinder. For 
\begin{figure}[ht]
{\centering \resizebox*{7.5cm}{9.5cm}{\includegraphics{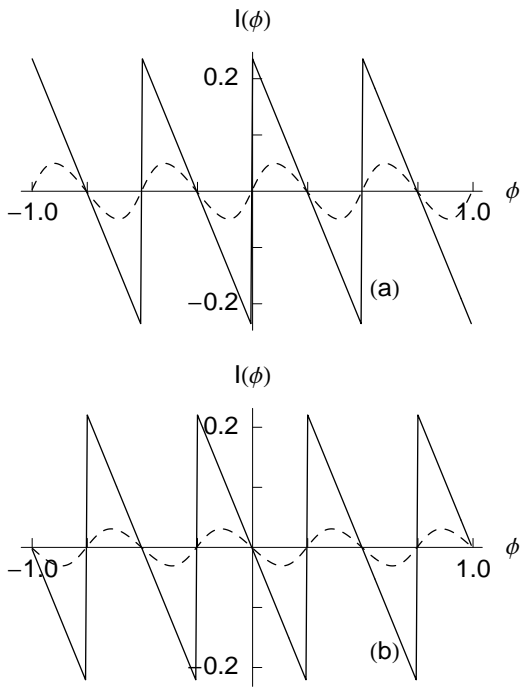}}\par}
\caption{{\textsl{$I$-$\phi$ characteristics for ordered (solid curve)
and disordered (dotted curve) moebius strips ($N=50$) with $v_{\bot}=0$.
(a) $N_e=40$ and (b) $N_e=35$. For the disordered case, $W=1$.}}}
\label{mobcurr1}
\end{figure}
illustrations, see the results plotted in Fig.~\ref{mobeng2}, where
the solid and dotted lines correspond to the identical meaning as
in Fig.~\ref{mobeng1}.

With the above energy-flux characteristics now we concentrate our study
on the current-flux spectra. Let us begin with the results plotted in
Fig.~\ref{mobcurr1}, where the currents are evaluated for some moebius
strips with $v_{\bot}=0$. The system size $N$ is fixed at $50$, where
(a) and (b) correspond to $N_e=40$ (even) and $35$ (odd), respectively.
The solid and dotted curves represent the currents for ordered and 
disordered moebius strips, respectively, where the disorder strength
$W$ is set to $1$. Like single channel mesoscopic ring, persistent 
current reveals saw-tooth like behavior as a function of $\phi$ both
for odd and even number of electrons. Similarly, in presence of disorder
\begin{figure}[ht]
{\centering \resizebox*{7.5cm}{9.5cm}{\includegraphics{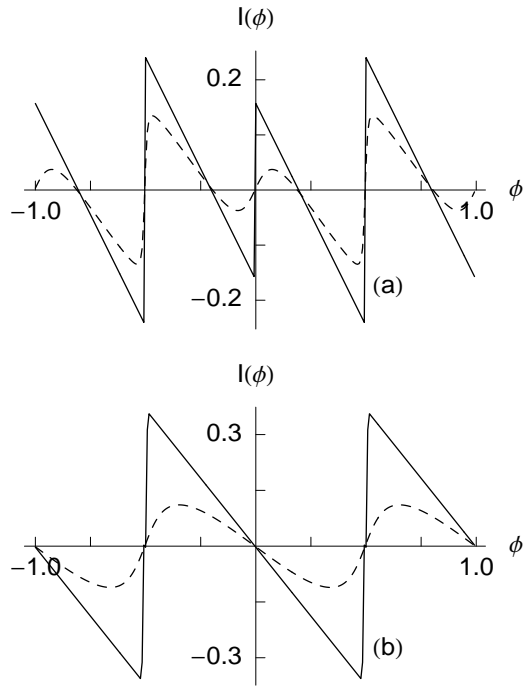}}\par}
\caption{{\textsl{$I$-$\phi$ characteristics for ordered (solid curve)
and disordered (dotted curve) moebius strips ($N=50$) with $v_{\bot}=-1$.
(a) $N_e=40$ and (b) $N_e=35$. For the disordered case, $W$ is fixed at 
$1$.}}}
\label{mobcurr2}
\end{figure}
current varies continuously, as studied earlier, and gets much reduced
value than the perfect case. These features are clearly understood from
our previous discussion. The main point is that, for this particular
case where $v_{\bot}=0$, current varies periodically with $\phi$ showing
$\phi_0/2$ periodicity, instead of conventional $\phi_0$ periodicity.
This unconventional period halving behavior of persistent current may
be clearly visible from our energy-flux spectrum, for instance see
Fig.~\ref{mobeng1}. 

The situation becomes quite different when electrons can able to hop
along the transverse direction i.e., for $v_{\bot} \ne 0$. In this case
a moebius can be regarded as an ordinary mesoscopic cylinder, and
therefore, we get the conventional features of persistent current.
For illustrative purposes, in Fig.~\ref{mobcurr2} we show the 
current-flux characteristics for some typical moebius strips ($N=50$)
considering $v_{\bot}=-1$, where (a) and (b) correspond to the 
results for $N_e=40$ and $35$, respectively. The solid and dotted
lines represent the similar meaning as in Fig.~\ref{mobcurr1}.
The saw-tooth like behavior is still preserved in perfect case. But, 
some additional kinks may appear in $I$-$\phi$ spectrum at different 
values of $\phi$ depending on the choices of $N_e$. For instance,
the current shows a kink across $\phi=0$ when $N_e=40$. These kinks
are associated with the crossing of energy levels for the cylindrical
systems. While, for disordered cases, current varies continuously
with much reduced amplitude, as expected, and no kink will appear. 

\section{{\textsl{Low-field magnetic susceptibility}}}

Next, we focus our attention on the determination of magnetic 
susceptibility $\chi(\phi)$ in the limit $\phi\rightarrow 0$. 
Evaluating the sign of $\chi(\phi)$ one can able to predict whether the
current is paramagnetic or diamagnetic in nature. It was a long standing
problem over the past few many years, and, here we try to emphasize about 
it both for one-channel rings and moebius strips described with fixed
number of electrons $N_e$. The general expression of magnetic 
susceptibility at any flux $\phi$ is expressed in the form,
\begin{equation}
\chi(\phi)=\frac{N^3}{16\pi^2}\left(\frac{\partial I(\phi)}{\partial \phi}
\right)
\end{equation}
where, $N$ corresponds to the total number of atomic sites in the system.
Here we will determine $\chi(\phi)$ only in the limit $\phi \rightarrow 0$
i.e., we are interested only on the low-field magnetic susceptibility.

\begin{figure}[ht]
{\centering \resizebox*{7.5cm}{9.5cm}{\includegraphics{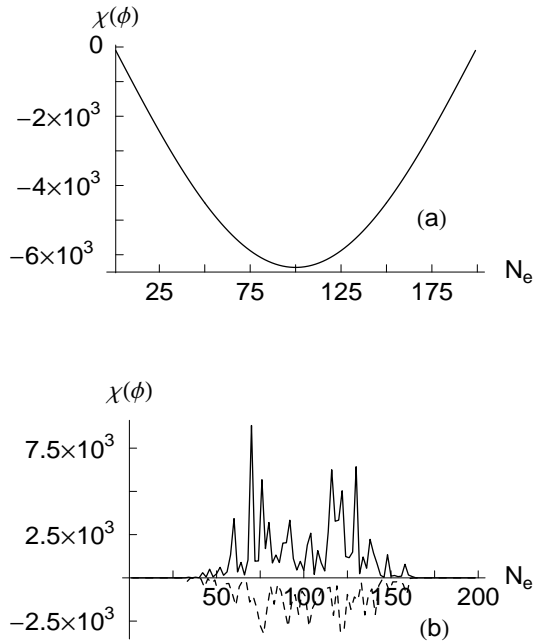}}\par}
\caption{{\textsl{Low-field magnetic susceptibility $\chi(\phi)$ as a
function of $N_e$ for (a) ordered rings and (b) disordered ($W=1$) rings
considering $N=200$. The solid and dotted curves in (b) correspond to the
results for even and odd $N_e$, respectively.}}}
\label{ringsus}
\end{figure}

As representative examples, in Fig.~\ref{ringsus} we display the 
variation of low-field magnetic susceptibility as a function of $N_e$
for strictly one-channel mesoscopic rings, where (a) and (b) correspond
to the perfect ($W=0$) and disordered ($W=1$) rings, respectively. The 
system size $N$ is fixed at $200$. Our results predict that in 
absence of any impurity $\chi(\phi)$ is always negative irrespective
of the total number of electrons $N_e$ i.e., whether $N_e$ is odd
or even. It reveals that for perfect rings low-field current provides 
only the diamagnetic response. This diamagnetic behavior of low-field
currents can be clearly understood by noting the slope of $I$-$\phi$
characteristics plotted by the solid lines in Figs.~\ref{ringcurr}(a) 
and (b). The response drastically changes as long as impurities are 
introduced in the system (\ref{ringsus}(b)). For odd and even $N_e$,
it shows completely opposite behavior. The low-field current gives
paramagnetic response for the rings with even $N_e$ (solid line), 
while for the rings with odd $N_e$ it shows diamagnetic nature (dotted 
line). These diamagnetic and paramagnetic natures of persistent current 
for the disordered case can be easily understood by observing the slopes of 
$I$-$\phi$ curves presented by the dotted lines in Figs.~\ref{ringcurr}(a) 
and (b). All these features are equally valid irrespective of disorder 
strength and disordered configurations. Thus, in brief we can say that
for one-channel rings described with fixed number of electrons, sign 
of low-field currents can be precisely determined.

At the end, we focus on the behavior of low-field magnetic response 
in moebius geometries. For these systems, sign of low-field currents 
strongly depends on the strength of intra-chain interaction ($v_{\bot}$). 
For $v_{\bot}=0$, a moebius strip becomes a single channel ring, and
therefore, sign of low-field currents can be mentioned exactly according
to the above prescription. On the other hand for the case where $v_{\bot}$
is finite, sign of low-field currents cannot be predicted accurately as 
it strongly depends on the number of electrons and specific realization 
of disordered configuration in the strips.

\section{{\textsl{Concluding remarks}}}

To summarize, we have studied the behavior of persistent current and
low-field magnetic susceptibility in strictly one-channel mesoscopic
rings and moebius strips penetrated by magnetic flux $\phi$. A 
tight-binding framework is given to describe the system, where all
calculations are done numerically to study the magnetic response. In 
perfect rings current shows saw-tooth like variation as a function of 
$\phi$, while it varies continuously in disordered rings. Both for 
perfect and disordered rings current varies periodically with $\phi$ 
exhibiting $\phi_0$ flux-quantum periodicity. In moebius strips a 
unconventional $\phi_0/2$ periodic nature of persistent current is 
observed for the particular case when $v_{bot}=0$. While, current 
gets back its $\phi_0$ periodicity when electrons are able to hop 
along the transverse direction i.e., $v_{\bot} \ne 0$. In the study 
of low-field magnetic response we have seen that for one-channel
mesoscopic rings described with constant number of electrons, sign 
of low-field currents can be mentioned precisely. While, for moebius
strips with $v_{\bot} \ne 0$, sign of low-field currents cannot be
predicted exactly as it depends on the number of electrons as well
as disordered configurations. Our exact analysis can provide some
physical insight to study magnetic response in mesoscopic loop
geometries.


\begin{thebibliography}{99}

\bibitem{byers} N. Byers and C. N. Yang, Phys. Rev. Lett. \textbf{7},
46 (1961).
\bibitem{butt} M. B\"{u}ttiker, Y. Imry, and R. Landauer, Phys. Lett.
\textbf{96A}, 365 (1983).
\bibitem{cheu1} H. F. Cheung, Y. Gefen, E. K. Riedel, and W. H. Shih,
Phys. Rev. B \textbf {37}, 6050 (1988).
\bibitem{cheu2} H. F. Cheung and E. K. Riedel, Phys. Rev. Lett.
\textbf{62}, 587 (1989).
\bibitem{mont} G. Montambaux, H. Bouchiat, D. Sigeti, and R. Friesner,
Phys. Rev. B \textbf{42}, 7647 (1990).
\bibitem{alts} B. L. Altshuler, Y. Gefen, and Y. Imry, Phys. Rev. Lett.
\textbf{66}, 88 (1991).
\bibitem{von} F. von Oppen and E. K. Riedel, Phys. Rev. Lett.
\textbf{66}, 84 (1991).
\bibitem{schm} A. Schmid, Phys. Rev. Lett. \textbf{66}, 80 (1991).
\bibitem{ambe} V. Ambegaokar and U. Eckern, Phys. Rev. Lett.
\textbf{65}, 381 (1990).
\bibitem{bouz} G. Bouzerar, D. Poilblanc, and G. Montambaux, Phys.
Rev. B \textbf{49}, 8258 (1994).
\bibitem{giam} T. Giamarchi and B. S. Shastry, Phys. Rev. B \textbf{51},
10915 (1995).
\bibitem{yu} N. Yu and M. Fowler, Phys. Rev. B \textbf{45}, 11795 (1992).
\bibitem{kulik} I. O. Kulik, Physica B \textbf{284}, 1880 (2000).
\bibitem{avishai} K. Yakubo, Y. Avishai, and D. Cohen, Phys. Rev. B
\textbf{67}, 125319 (2003).
\bibitem{weiden} E. H. M. Ferreira, M. C. Nemes, M. D. Sampaio, and
H. A. Weidenm\"{u}ller, Phys. Lett. A \textbf{333}, 146 (2004).
\bibitem{san1} S. K. Maiti, Int. J. Mod. Phys. B \textbf{21}, 179 (2007).
\bibitem{san2} S. K. Maiti, J. Chowdhury, and S. N. Karmakar, Synthetic
Metals \textbf{155}, 430 (2005).
\bibitem{san3} S. K. Maiti, J. Chowdhury, and S. N. Karmakar,
J. Phys.: Condens. Matter \textbf{18}, 5349 (2006).
\bibitem{levy} L. P. Levy, G. Dolan, J. Dunsmuir, and H. Bouchiat,
Phys. Rev. Lett. \textbf{64}, 2074 (1990).
\bibitem{chand} V. Chandrasekhar, R. A. Webb, M. J. Brady, M. B. Ketchen,
W. J. Gallagher, and A. Kleinsasser, Phys. Rev. Lett. \textbf{67},
3578 (1991).
\bibitem{jari} E. M. Q. Jariwala, P. Mohanty, M. B. Ketchen, and R. A. Webb,
Phys. Rev. Lett. \textbf{86}, 1594 (2001).
\bibitem{deb} R. Deblock, R. Bel, B. Reulet, H. Bouchiat, and D. Mailly,
Phys. Rev. Lett. \textbf{89}, 206803 (2002).
\bibitem{san4} S. K. Maiti, Phy. Scr. \textbf{73}, 519 (2006);
[Addendum: Phy. Scr. \textbf{78}, 019801 (2008)].
\bibitem{san5} S. K. Maiti, Physica E \textbf{31}, 117 (2006).
\bibitem{lee} P. A. Lee and T. V. Ramakrishnan, Rev. Mod. Phys. 
\textbf{57}, 287 (1985).

\end{thebibliography}
\end{document}